\documentclass[11pt,a4paper]{article}
\usepackage{jcappub}
\usepackage{bm}
\usepackage{bbm}
\usepackage{amsmath}
\usepackage{etoolbox}
\usepackage{float}
\usepackage{mhchem}
\usepackage{gensymb}
\usepackage{color}
\usepackage{appendix}
\usepackage{hyperref}

\definecolor{seagreen}{rgb}{0.18, 0.55, 0.34}
\definecolor{forestgreen}{rgb}{0.13, 0.55, 0.13}
\usepackage{hyperref}
\hypersetup{
    colorlinks=true,
    linkcolor=forestgreen,      
    urlcolor=forestgreen,       
    filecolor=forestgreen,
    citecolor=forestgreen,      %
    linktoc=page 
}

\def\circa#1{\,\raise.3ex\hbox{$#1$\kern-.75em\lower1ex\hbox{$\sim$}}\,}

\newcommand{\DM}{{\rm DM}}

\newcommand{\beq}{\begin{equation}}
\newcommand{\eeq}{\end{equation}}

\definecolor{seagreen}{rgb}{0.18, 0.55, 0.34}

\usepackage{etoolbox}

\makeatletter
\gdef\@fpheader{}
\makeatother

\begin{document}

\makeatletter

\title{Searching for dark matter X-ray lines from the Large Magellanic Cloud with eROSITA}

\author[a]{Jorge Terol Calvo,}
\author[a]{Marco Taoso,}
\author[b,c,d]{Andrea Caputo,}
\author[e]{\\Michela Negro,}
\author[f,a]{Marco Regis}

\affiliation[a]{Istituto Nazionale di Fisica Nucleare, Sezione di Torino, via P. Giuria 1, I--10125 Torino, Italy}
\affiliation[b]{Department of Theoretical Physics, CERN, Esplanade des Particules 1, P.O. Box 1211, Geneva 23, Switzerland}
\affiliation[c]{Dipartimento di Fisica, ``Sapienza'' Universit\`a di Roma \& Sezione INFN Roma1, Piazzale Aldo Moro
5, 00185, Roma, Italy}
\affiliation[d]{Department of Particle Physics and Astrophysics, Weizmann Institute of Science, Rehovot 7610001, Israel}
\affiliation[e]{Department of Physics and Astronomy, Louisiana State University, Baton Rouge, LA 70803, USA}
\affiliation[f]{Dipartimento di Fisica, Universit\`{a} di Torino, via P. Giuria 1, I--10125 Torino, Italy}

\emailAdd{terolcal@to.infn.it}
\emailAdd{marco.taoso@to.infn.it}
\emailAdd{andrea.caputo@cern.ch}
\emailAdd{michelanegro@lsu.edu}
\emailAdd{marco.regis@unito.it}

\abstract{We perform a search for an X-ray monochromatic line arising from dark matter (DM) decay in the halo of the Large Magellanic Cloud. An emission line can be expected from two well-motivated DM candidates: sterile neturinos and axion-like particles (ALPs). We analyze the eROSITA-DE DR1 datasets in the energy range between 1 and 9 keV. No evidence for a DM line is found, and we set lower limits on the DM lifetime. We then recast these bounds into upper limits on the active-sterile neutrino mixing angle $\sin^2(2\theta)$ and on the ALP to photon coupling $g_{a\gamma}$, for DM masses between 2 and 18 keV. 
These results set new strong constraints for masses below 5 keV.}

\maketitle
\section{Introduction}\label{Sec:Introduction}
Identifying the nature of dark matter (DM) remains one of the central challenges in modern physics. Although its gravitational effects are firmly established across a wide range of astrophysical and cosmological observations, its particle nature and properties are still unknown. 

Among the well-motivated particle candidates beyond the Standard Model are sterile neutrinos in the keV mass range~\cite{Dodelson:1993je,Asaka:2005pn,Abazajian:2017tcc} and axion-like particles (ALPs)~\cite{Arias:2012az,Ringwald:2012hr, Foster:2022ajl, Agrawal:2025rbr, Reig:2025dqb}, including the Majoron~\cite{Berezinsky:1993fm, Lattanzi:2007ux, Bazzocchi:2008fh, Queiroz:2014yna, Garcia-Cely:2017oco}. Both classes of candidates naturally arise in extensions of the Standard Model, can account for the observed DM abundance in well-defined regions of parameter space, and predict observable signatures in the X-ray band.

One of their distinctive features is the possibility of radiative decay into photons. In particular, sterile neutrinos with mass $m_s$ in the keV range can undergo radiative  decay into an active neutrino and a monoenergetic photon, $\nu_s\rightarrow \nu_a+\gamma$, producing a narrow X-ray line at an energy $E_\gamma \simeq m_s/2$
with decay rate~\cite{Lee:1977tib, Pal:1981rm}: 
\begin{equation}
\label{eq:rateSterile}
    \Gamma_{\nu_s\rightarrow \nu_a \gamma} \sim (7.2\,\, 10^{29}{\rm s})^{-1}\Big(\frac{\sin^2(2\theta)}{10^{-8}}\Big) \Big(\frac{m_s}{1\,{\rm keV}}\Big)^5\;,
\end{equation}
where $\theta$ is the mixing angle between active and sterile neutrinos. Similarly, ALPs of mass $m_a$ can decay into two photons, $a\rightarrow \gamma \gamma$, with rate
\begin{equation}
\label{eq:rateALP}
\Gamma_{a \to \gamma\gamma} = \frac{g_{a\gamma}^2 m_a^3}{64\pi},
\end{equation}
where $g_{a\gamma}$ denotes the ALP–photon coupling. 

In both cases, the expected photon flux from a given target scales with the DM column density (the so-called $D$-factor), i.e., it is linearly proportional to the integral of the DM density along the line-of-sight.
X-ray observations of DM-dominated objects can be a powerful probe for sterile neutrino and ALP DM.

Over the past decade, extensive searches for narrow X-ray lines have been conducted in galaxies, galaxy clusters, and the Milky Way halo~\cite{Boyarsky:2005us,Boyarsky:2006zi,Boyarsky:2007ge,Watson:2006qb,Yuksel:2007xh,Loewenstein:2008yi,Riemer-Sorensen:2009zil,Boyarsky:2008ju,Horiuchi:2013noa,Urban:2014yda,Anderson:2014tza,Tamura:2014mta,XQC:2015mwy,Iakubovskyi:2015dna,Ng:2015gfa, Sekiya:2015jsa,Riemer-Sorensen:2015kqa,Ruchayskiy:2015onc,Neronov:2016wdd,Perez:2016tcq,Ng:2019gch,Roach:2019ctw,
Foster:2021ngm,Roach:2022lgo,Krivonos:2024yvm,XRISM:2025lzv}, including investigations of the debated 3.5 keV line~\cite{Bulbul:2014sua,Boyarsky:2014jta,Dessert:2018qih,Boyarsky:2020hqb,Dessert:2023fen}.  While no conclusive evidence for decaying DM has emerged, these studies have placed stringent bounds on the sterile neutrino mixing angle and on the ALP parameter space across the 1-100 keV mass range. More recently, complementary constraints have been derived from diffuse X-ray observations with eROSITA~\cite{Balaji:2025afr} and from the ionization of molecular clouds~\cite{DelaTorreLuque:2025zjt}.

The Galactic Center (GC) and the Milky Way halo are ideal targets due to their high expected DM signal, although searches require a careful treatment of the intense and crowded astrophysical backgrounds in the GC or the modeling of very large spatial regions.
The Large Magellanic Cloud (LMC) represents a particularly promising alternative.
As the most massive satellite of the Milky Way, located at a distance of $\sim 50$ kpc~\cite{Pietrzynski:2019cuz}, the LMC combines a substantial DM halo with close proximity, resulting in a sizable decay flux, i.e., a large $D$-factor.
Compared to the Milky Way halo, its smaller angular extent makes large-scale analysis more manageable, and although it hosts strong X-ray-emitting components, such as hot interstellar gas, supernova remnants and X-ray binaries, the environment is less crowded and complex than the GC, simplifying spectral modeling.

The extended ROentgen Survey with an Imaging Telescope
Array (eROSITA) onboard the observatory satellite Spectrum-Roentgen-Gamma (SRG)~\cite{Merloni:2012uf,Predehl:2021aqx} provides an unprecedented opportunity to search for faint, spatially extended X-ray signals. 
The grasp of eROSITA mission,
defined as the product of field of view times and effective area, surpasses those of current and previous X-ray telescopes in the 0.3-3.5 keV energy range, while remaining competitive at higher energies~\cite{eROSITATechnical}.
The energy resolution, expressed as the full width at half maximum (FWHM), ranges from approximately 70 to 170 eV over the 1–10 keV energy range~\cite{eROSITATechnical}. The large grasp, its good spectral resolution, and survey strategy initiated in December 2019 make eROSITA particularly well suited for searches for faint, spatially extended line emission from nearby systems.

The eROSITA-DE first data release (DR1)~\footnote{Details of the eROSITA-DE Data Release 1 (DR1) are available at~\href{https://erosita.mpe.mpg.de/dr1/}{https://erosita.mpe.mpg.de/dr1/}} comprises data from the first six months of observations, and covers 20626.5 square degrees of the sky in the 0.2-10 keV energy band~\cite{2024A&A...682A..34M}.

In this work, we analyze eROSITA DR1 observations of the LMC to search for monochromatic X-ray lines consistent with decaying DM. We perform a spectral analysis accounting for astrophysical and instrumental backgrounds to set upper limits on line emission in the 1-9 keV energy range. We then translate these results into constraints on sterile neutrino DM in the $(m_s,\,\sin^2(2 \theta))$ plane and on ALP DM in the $(m_a,\,g_{a\gamma})$ plane. 
Our results provide competitive and complementary limits with respect to previous X-ray searches, especially for DM masses below 5 keV.

The paper is organized as follows: Section~\ref{Sec:Data}
describes the eROSITA data and the procedure used to construct spectra of the LMC.
Section~\ref{Sec:methodology} presents the methodology for fitting the data and accounting for the various expected contributions, including instrumental, astrophysical, and DM components.
In Section~\ref{Sec:Results}, we report the results in terms of bounds on sterile neutrino and ALP DM.
Section~\ref{Sec:Conclusion} concludes.

\begin{figure}[H]
    \centering
    \includegraphics[width=0.75\textwidth]{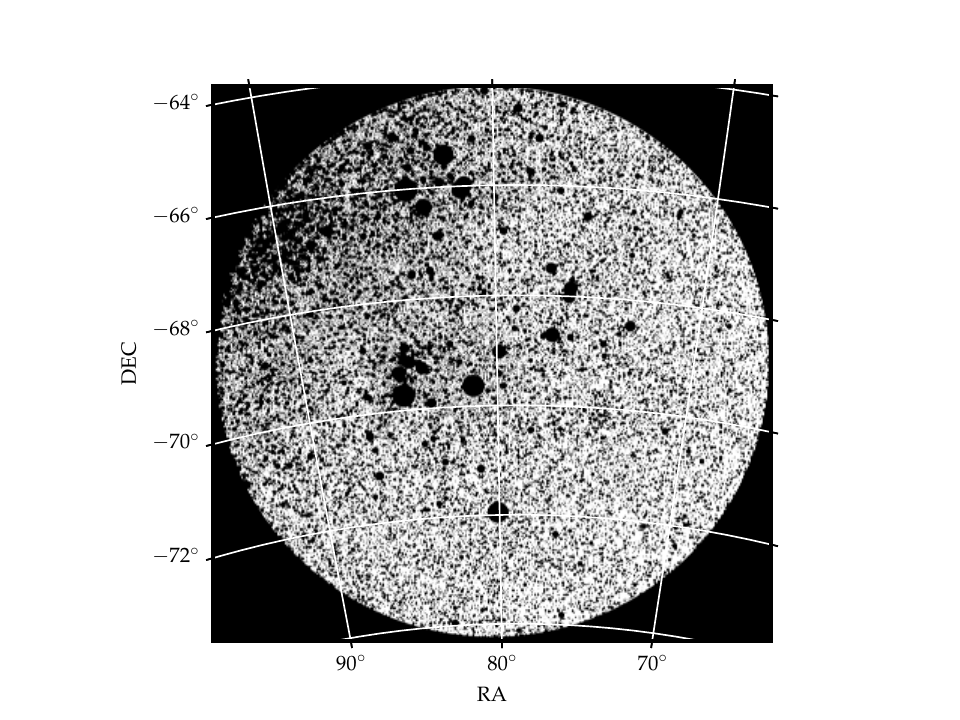}
    \centering
    \caption{Source-exclusion mask (``cheesemask'') used for the analysis. Pixels shown in black are excluded from the spectral extraction.}
    \label{fig:CheeseMask}
\end{figure}

\section{Data}
\label{Sec:Data}

\begin{figure}
\centering
\includegraphics[width=0.75\textwidth]{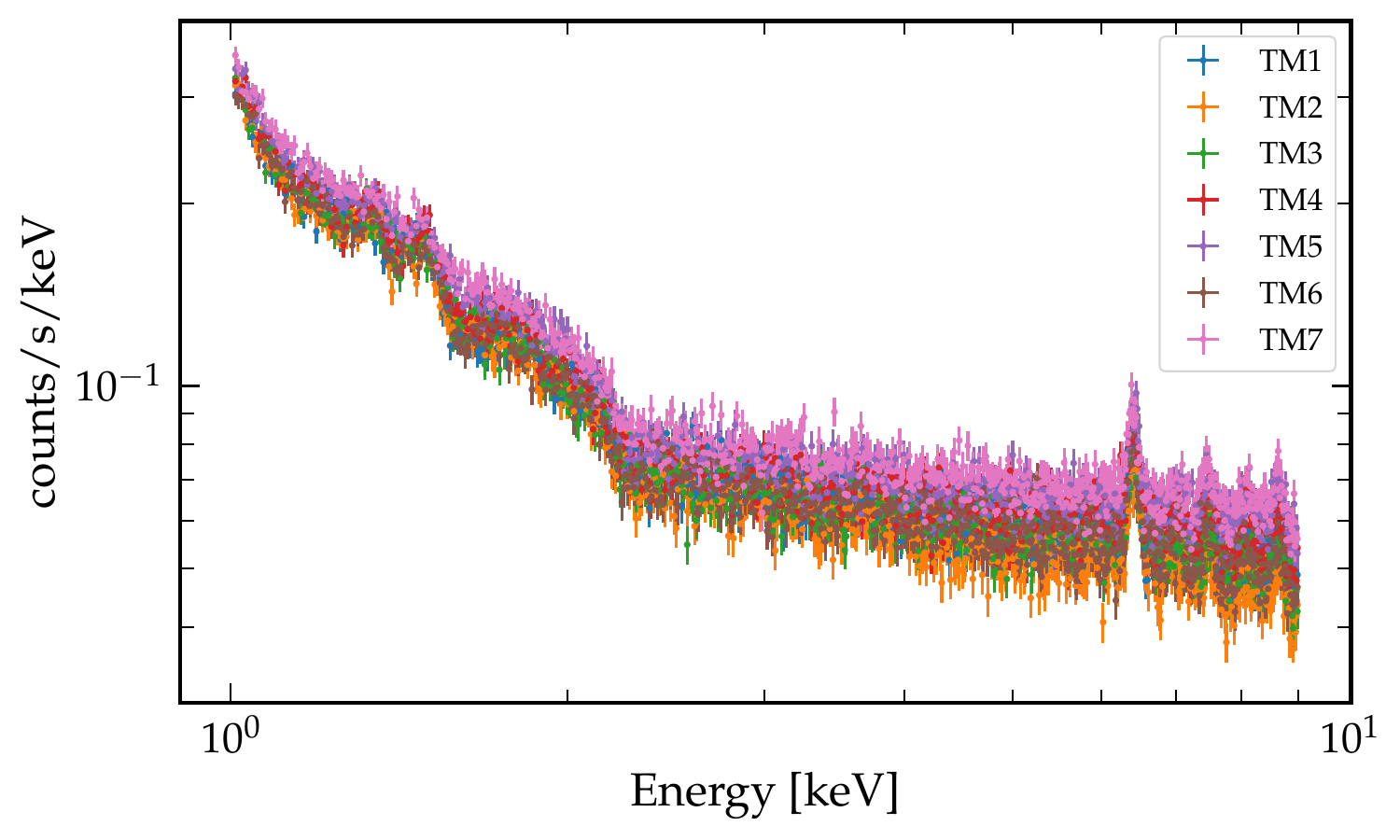}
\centering
\caption{Count rate in the 1–9 keV energy range for all seven TMs within the extraction region defined by the cheesemask in Fig.~\ref{fig:CheeseMask}. See text for details.}
\label{fig:Total_RateSpectrum}
\end{figure}

We extracted counts in a region of $5\degree$ radius around the kinematic center of the LMC, taken to be at $\rm{RA}=80.26\degree$, $\rm{DEC} = -69.26\degree$ from~\cite{2024A&A...692A..40K} and based on Jeans modeling of the stellar kinematics.
The radius covers the spatial extent of the kinematic tracers used to infer the DM density distribution in~\cite{2024A&A...692A..40K}. We started downloading DR1 data~\cite{2024A&A...682A..34M} corresponding to the sky tiles intersecting such region. Then, we used the eSASS~\cite{2022A&A...661A...1B} v4 DR1 software\footnote{\href{https://erosita.mpe.mpg.de/dr1/eSASS4DR1/}{https://erosita.mpe.mpg.de/dr1/eSASS4DR1/}} to process the data and extract the spectrum. 
First, we combined all the counts of the different sky tiles, using the \texttt{evtool} task, and filtered out the ones that are not enclosed in the 5 deg radius circular region around the center of the LMC. To create images suitable for source detection and spectral extraction, we generated a binned, energy-filtered count map between 0.5 and 10 keV over the full region of interest, again with \texttt{evtool}. The image was rebinned with a bin size of 80 (detector pixels), without a significant loss of spatial resolution and still suitable for source masking. Accurate exposure maps accounting for spacecraft altitude, instrumental vignetting, and CCD live time are required for both source detection and spectral analysis. We computed a merged exposure map matched to the count image using the \texttt{expmap} task. We executed a detection run with \texttt{erbox} to generate a list of candidate excesses using a conservative detection threshold $\mathit{Detection\ likelihood>6}$ as our aim is not source cataloguing but building a conservative exclusion mask for the diffuse spectrum. The following step was to produce a ``cheesemask'' with the \texttt{erbackmap} task and using the previously generated image, exposure map and source list. Circular regions around the input sources were calculated where the surface brightness of the input sources exceeded the threshold, set to the default value of $10^{-4}$ cts/px. 
This cheesemask excludes bright sources and artifacts, enabling diffuse emission and extended spectra to be extracted with minimal contamination. We set a conservative 70\% as the maximum allowed fraction of masked out area, finally getting masked $\sim 38\%$ of the area. The obtained mask is shown in Fig.~\ref{fig:CheeseMask}. 
Finally, we extracted spectra and instrumental response files using the \texttt{srctool} task, with the cheesemask defining the extraction region. 
eROSITA consists of seven independent mirror modules (TMs). As a result, for each TM, we are left with the spectrum file for the unmasked region, and the associated Ancillary Response File (ARF) and Redistribution Matrix File (RMF), which account for the effective area and energy response of the instrument, respectively.
The count rate is shown in Fig.~\ref{fig:Total_RateSpectrum} for all seven TMs in the energy range 1-9 keV, in which we performed the line search, as explained in the next section.

\section{Line search: methodology}
\label{Sec:methodology}

We perform the search for a DM line signal in the energy range 1-9 keV,
 neglecting the band below 1 keV because it is strongly contaminated by astrophysical absorption and emission lines, which makes the search for an exotic DM line extremely challenging, and avoiding data close to the upper edge (10 keV) where the count rate drops quickly. 

We describe the observed emission as the combination of the DM line signal and a background model, which includes the contribution from both astrophysical emission and the instrumental background. 
We will discuss in detail the model that we have implemented below. 
The data is analyzed considering Poisson statistics, and more specifically, computing the following log-likelihood:
\begin{equation}
C^k (\phi_{\rm DM},{ \vec \Pi})= 2\sum_i\Big[ \mu^k_i - n^k_i+n^k_i\,\log(n^k_i/\mu^k_i)\Big],~\,
\label{eq:likelihood}    
\end{equation}
where $n^k_i$ are the measured counts in the energy bin $i$ and in the TM $k$, while $\mu^k_i$ are the corresponding counts predicted by the model, which depend on the flux of the DM line signal $\phi_{\rm DM}$ (in photons/cm$^2$/s), and other parameters, called collectively $\vec \Pi$, which define the background model.

Eq.~\ref{eq:likelihood} is a modification of the standard log-Poisson likelihood. It is obtained subtracting a term that depends only on the data. Therefore, this does not affect bounds and evidence, which are computed through differences of log-likelihood values. On the other hand, the modification makes Eq.~\ref{eq:likelihood} to asymptotically follow the $\chi^2$ distribution in the large-count limit~\cite{Kaastra:2017abk}, and this allows us to compute the associated p-value, and establish the goodness of fit.

We compute the log-likelihood in Eq.~\ref{eq:likelihood} using the software PyXspec 2.1.2/\textsc{XSPEC} 12.13.1~\cite{1996ASPC..101...17A} (the appropriate statistics is implemented with the option {\it cstat}). The model counts $\mu^k_i$ are computed by \textsc{XSPEC} folding the theoretical model through the instrument response functions, the ARF and RMF matrices that we obtained for each TM,  and specifically for our target region, through the procedure described in Sec.~\ref{Sec:Data}.

\begin{figure}[t]
    \centering
    \includegraphics[width=0.48\textwidth]{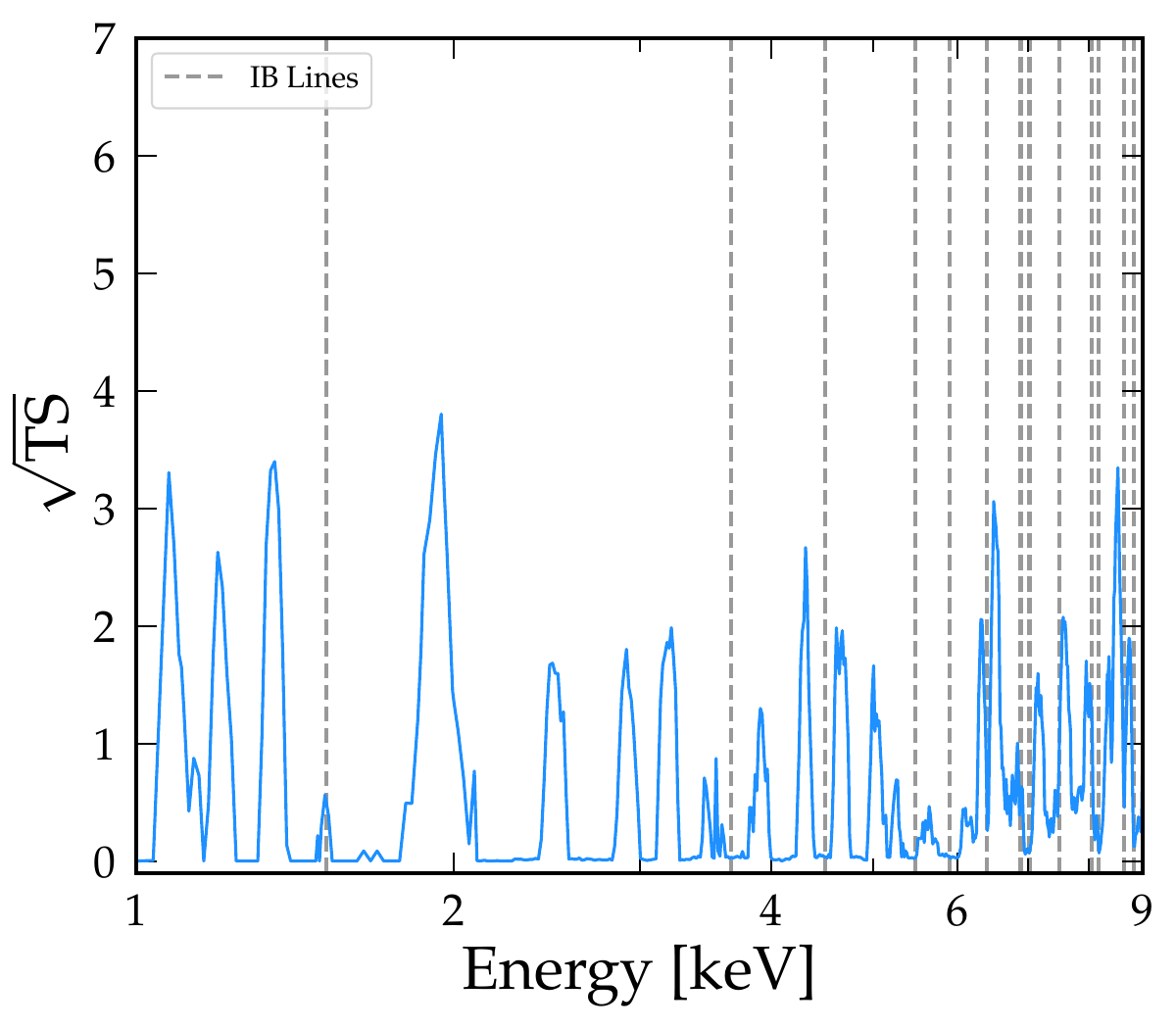}
    \includegraphics[width=0.48\textwidth]{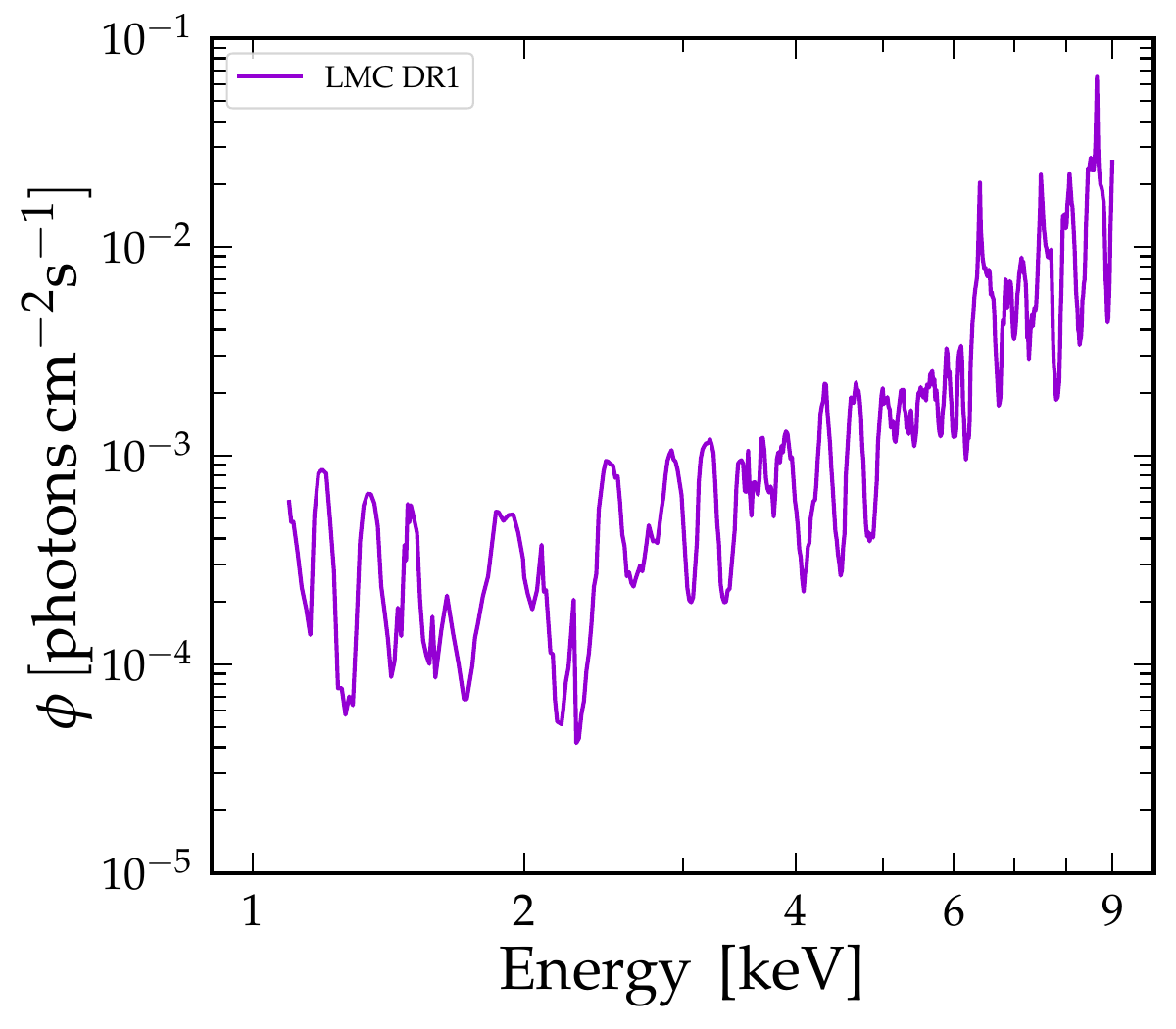}
    \centering
    \caption{\textit{Left:} Local significance $\sqrt{\rm TS}$ of line detection across the energy range studied. Dashed gray lines show the location of the instrumental background lines included in the analysis.  \textit{Right:} Upper limits on the the DM line flux across the energy range studied.}
    \label{fig:TS_scan}
\end{figure}

\begin{figure}[t]
   \centering
   \includegraphics[width=0.49\textwidth]{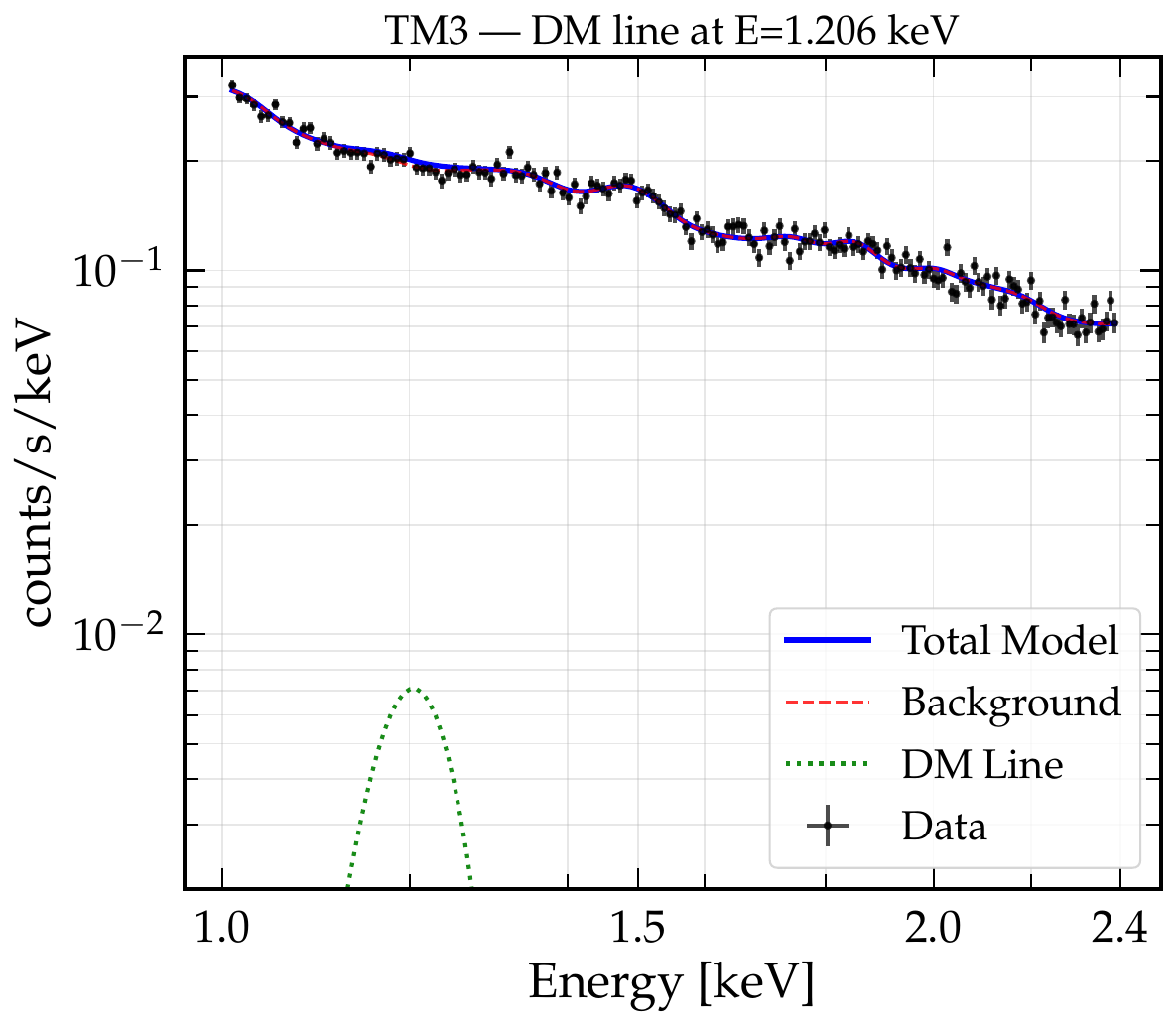}
   \includegraphics[width=0.49\textwidth]{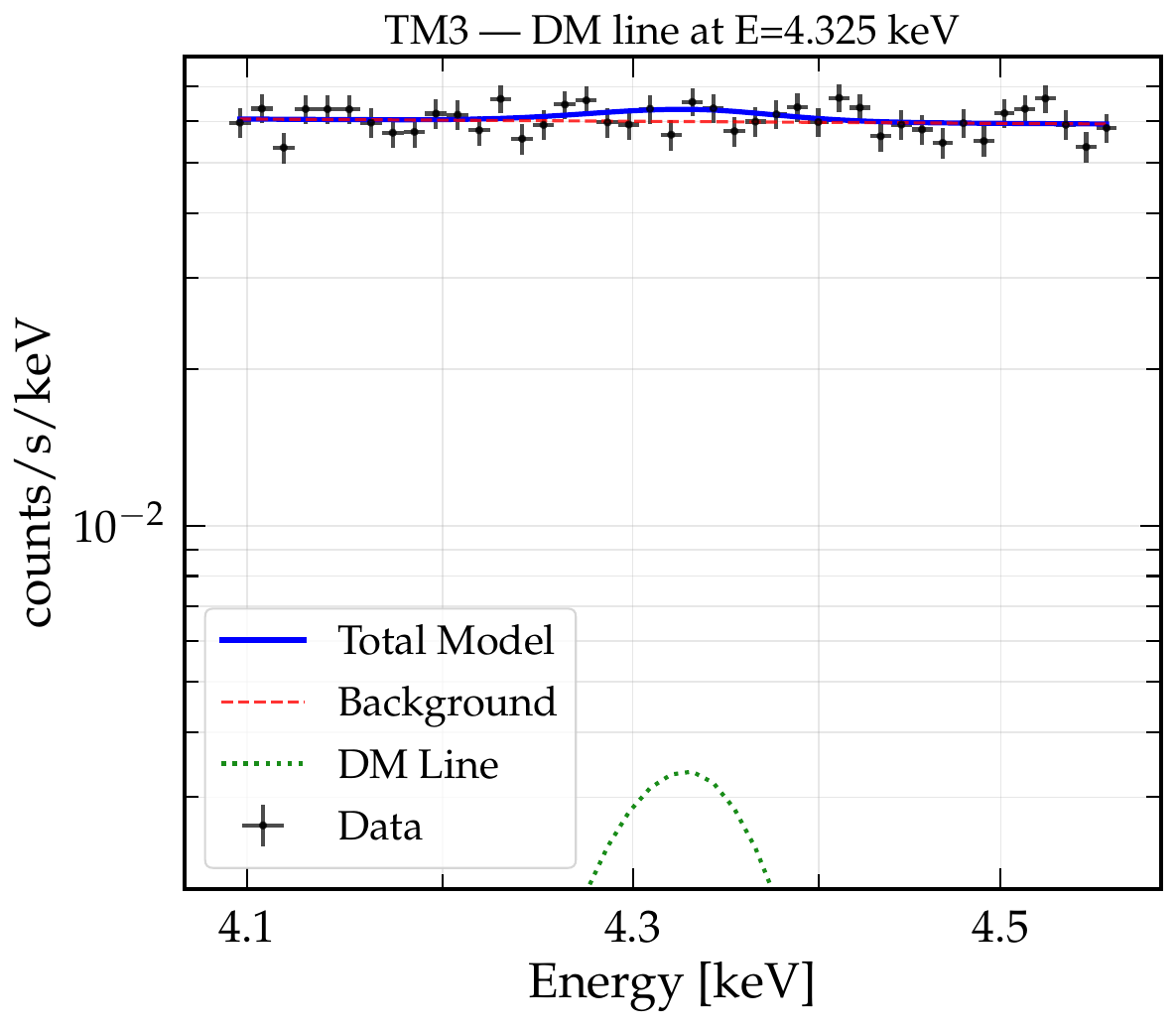}
   \centering
   \caption{Data and fitted spectra for two DM line searches in the eROSITA telescope module TM3. The red dashed line corresponds to the background model, the green dotted line to the DM signal corresponding to the 95\% C.L. upper limit (combining all seven TMs) and the blue solid line to their sum, i.e. the total fitted model. 
   Left and right panels refer to the search of a DM line at energies $E=1.206$ keV and $E=4.325$ keV, respectively.}
   \label{fig:DataWithModel}
\end{figure}
Finally, we construct the profile likelihood as

\begin{equation}
\tilde C^k (\phi_{\rm DM})= \rm{min}_{\vec \Pi} C^k (\phi_{\rm DM},{ \vec \Pi}),
\label{eq:proflikelihood}    
\end{equation}

obtained by minimizing the expression in Eq.~\ref{eq:likelihood} with respect to the nuisance parameters $\vec \Pi$ that define the background model. To search for the global minimum we combine the global optimizer {\it differential evolution} implemented in \textsc{Scipy}~\cite{2020SciPy-NMeth}, with local optimizer methods ({\it migrad} and {\it leven} implemented in \textsc{XSPEC}).

We now describe the components of the emission model. The DM line signal is implemented as a narrow gaussian, with a spread in energy related to the DM velocity dispersion in the LMC. Namely, we consider a width of the gaussian $\sigma=E_{\gamma}\,\sigma_v/c$ with $E_{\gamma}$ the photon energy and $\sigma_v\sim50$ km/s, although the precise value does not play any role since the width of the line is much smaller than the energy resolution of the instrument.  The possible absorption of the X-ray signal from the interstellar medium is modeled using the absorption model {\it TBABS} implemented in \textsc{XSPEC}. This component depends on the hydrogen column density $n_H,$ which we determine using the {\it HI4PI} map of 
HI\footnote{\href{https://lambda.gsfc.nasa.gov/product/foreground/fg_hi4pi_info.html}{https://lambda.gsfc.nasa.gov/product/foreground/fg\_hi4pi\_info.html}}. 
We find an average column density of $n_H=1.5\times 10^{21}\,{\rm cm}^{-2}$ over the unmasked analysis region.
We fix the $n_H$ parameter to this value in our analysis.

Using the 84th percentile of the HI4PI $n_H$ distribution within the same region of interest instead of the average value changes the inferred dark matter line flux limits by $\sim17\%$ at 1 keV, with a smaller impact at higher energies.
Absorption is negligible for energies $\gtrsim 2$ keV but becomes progressively more relevant at lower energies. 

As mentioned above, the background consists of both emission of astrophysical origin and an instrumental contribution. 
The instrumental background has been modeled from repeated eROSITA observations with  the filter-wheel set in CLOSED position~\cite{EDRFWC}. 
The eROSITA background model presented in~\cite{EDRFWC} is defined independently for each TM and consists of a linear combination of several narrow Gaussian lines and a smooth component modeled as the sum of a broken power law plus two additional power laws.
We find that the instrumental background dominates beyond $\sim2$ keV.
For this reason, we split the energy range into two bands, [1-2.3] keV and [2.3-9] keV, and analyze them separately. This approach allows a simpler and more conservative search for the DM line.

\smallskip

\begin{figure}[t]
   \centering
   \includegraphics[width=0.65\textwidth]{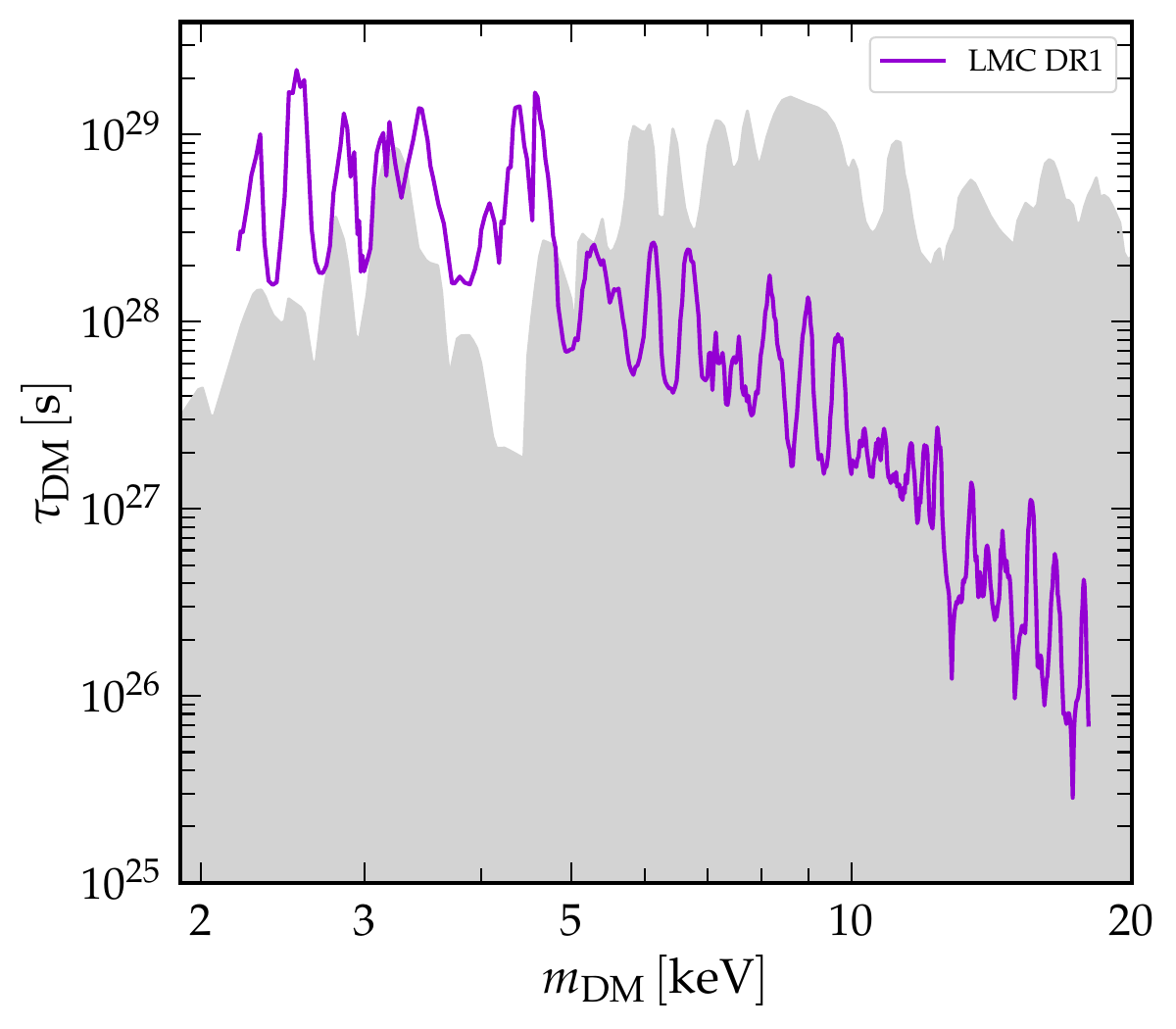}
   \centering
   \caption{Lower limits on the DM lifetime, $\tau_{\rm DM}$, as a function of the DM mass, $m_\DM$, for a two body decay process with one photon in the final state. In purple we show the limits derived in this paper using LMC observations from eROSITA DR1. In gray we show previously derived lower limits from other X-ray observations including NuSTAR~\cite{Roach:2022lgo, Krivonos:2024yvm}, XMM-Newton~\cite{Foster:2021ngm}, Chandra~\cite{Horiuchi:2013noa} and Suzaku~\cite{Loewenstein:2008yi}. 
   }
   \label{fig:ExclusionTau}
\end{figure}

\noindent {\bf Search for a DM line between 2.3 - 9 keV.} 
In this energy range the instrumental background dominates and we analyze the data using sliding energy windows around the putative DM line. Specifically, in Eq.~\ref{eq:likelihood}, we consider only the energy channels within a $\pm 5\sigma$ window centered on the energy of the DM line signal under consideration, where $\sigma$ indicates the energy resolution of each TM (FWHM = $2.355\sigma$)~\cite{eROSITATechnical}. 
We search for the DM line signal in energy steps of 15 eV, which is smaller than the energy resolution of eROSITA.
For these small windows and for each TM, we find that the background can be well approximated by a single power law, locally approximating the combination of power-laws of the global model adopted by eROSITA as mentioned above, plus a set of gaussian lines. 
The background is folded only through the RMF response function.

We consider the 12 instrumental lines reported in~\cite{EDRFWC} that fall within the 2.4–9 keV energy band.
Following~\cite{EDRFWC}, these lines are treated as intrinsically narrow, with the gaussian width set to $\sigma=0.1$ eV, much smaller than the eROSITA energy resolution. 
The energy of each instrumental line for each TM is kept fixed and determined as follows.
For each line and TM, we perform an initial fit in a window centered on the fiducial energy of the instrumental background model in~\cite{EDRFWC}. In this step, the line energy and the other background parameters are allowed to float, while the DM component is set to zero. We find the maximum of the likelihood in Eq.~\ref{eq:likelihood}, and record the best-fit line energies. They are then used as fixed parameters in the subsequent main analysis.
There, for each analyzed energy window, we include only the instrumental lines whose centroid energies lie within $\pm 7\sigma$ of the DM line.

Therefore, the free parameters of our background model, fitted to the data in Eq.~\ref{eq:proflikelihood}, are the normalizations of the relevant instrumental lines along with the normalization and spectral index of the power law. 

We emphasize that the parameters defining the instrumental background are specific to each TM; in other words, the fits in Eq.~\ref{eq:proflikelihood} are performed independently for each TM.
We implement this method and scan the energy of the DM line in steps smaller than the eROSITA energy resolution. For each step, we compute the profile likelihood in Eq.~\ref{eq:proflikelihood}.

The method we implement is robust and conservative with respect to the search for a signal line. Restricting the analysis to small energy windows, rather than fitting the entire energy range, minimizes potential background mismodeling that could otherwise produce spurious evidence for a line, see e.g. the discussion in~\cite{Dessert:2023fen}. Moreover, the reduced window size limits the number of background parameters, as only a small subset of instrumental lines needs to be included and the smooth emission can be well approximated by a single power law. 
This procedure simplifies the minimization of the statistic in Eq.~\ref{eq:proflikelihood}, improving the stability of the fit and the convergence toward the global minimum.
We also tested the stability of the line-search procedure against variations of the sliding-window width.
We find that defining the width of the sliding energy windows as, e.g., $\pm 4\sigma$ instead of $\pm 5\sigma$ affects the results only marginally: the statistical evidence for the DM line signal remains unchanged, and the corresponding limits on the signal flux vary only at the $20$\% level.

\begin{figure}[t]
   \centering
   \includegraphics[width=0.49\textwidth]{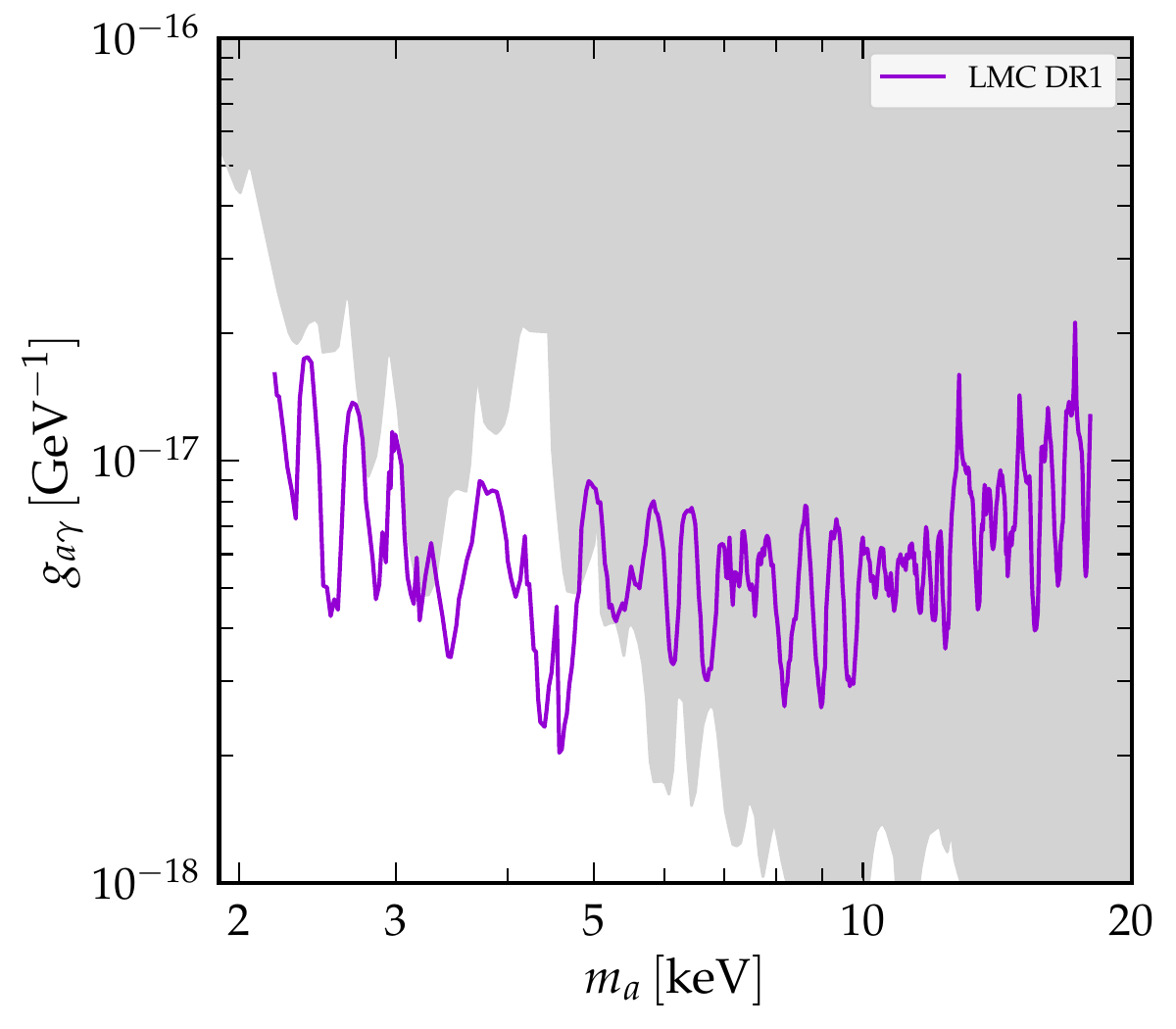}
   \includegraphics[width=0.49\textwidth]{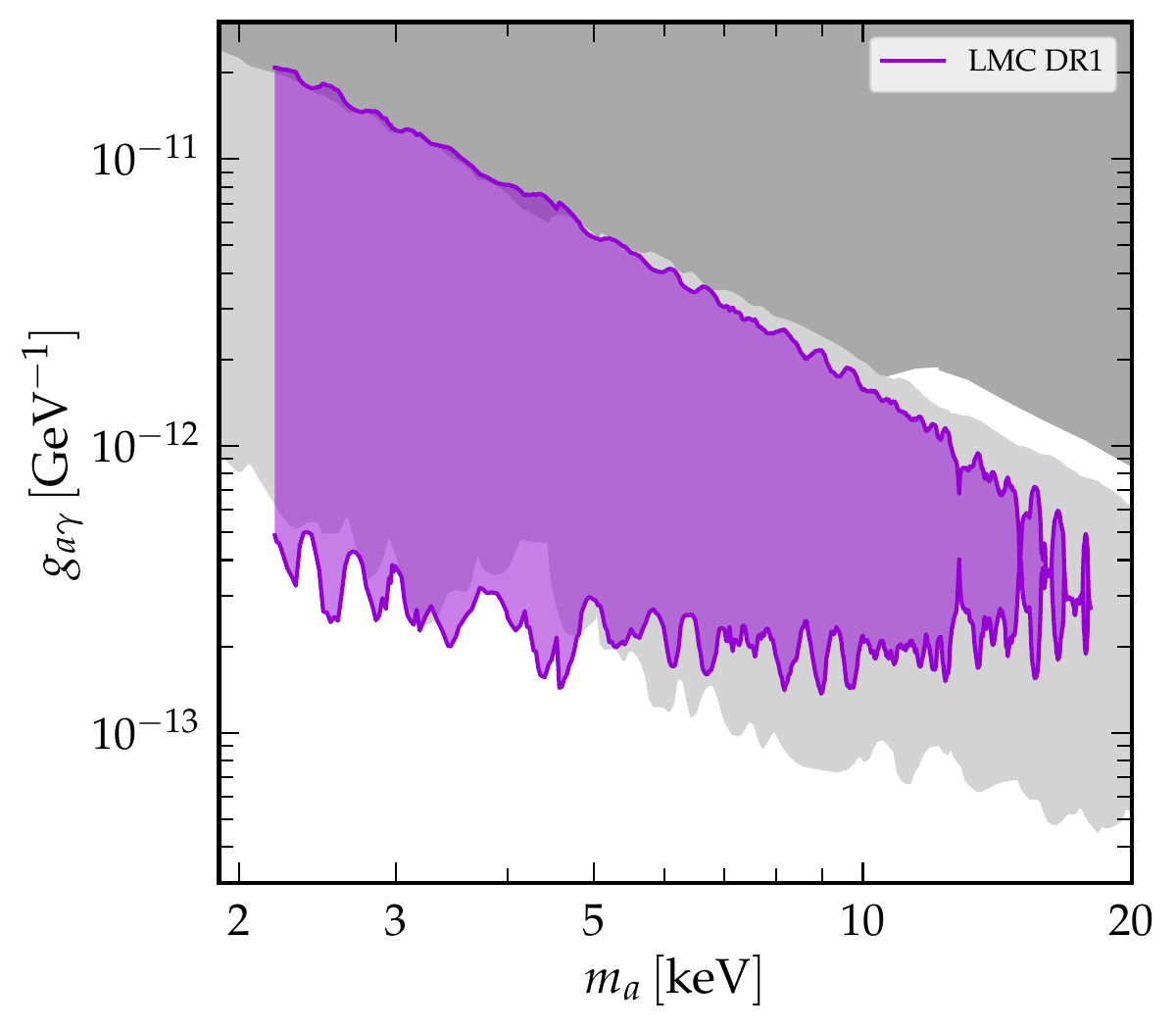}
   \centering
   \caption{Upper limits on the axion to photon coupling, $g_{a\gamma}$, as a function of the axion mass, $m_a$. On the left panel the limts are derived under the assumption that ALPs make up for the entire DM relic density while on the right panel constraints are derived only taking into account the irreducible axion background contribution~\cite{Langhoff:2022bij}, assuming $T_{\rm RH} = 5\,$MeV. In purple we show the limits derived in this work using LMC observations from eROSITA DR1. In light gray we show previously derived lower limits from other X-Ray observations including NuSTAR~\cite{Roach:2022lgo, Krivonos:2024yvm}, XMM-Newton~\cite{Foster:2021ngm}, Chandra~\cite{Horiuchi:2013noa} and Suzaku~\cite{Loewenstein:2008yi}. In the case of the irreducible axion background we also show in dark gray constraints coming from CRB~\cite{Hill:2018trh} and CMB~\cite{Balazs:2022tjl}.
   }
   \label{fig:ExclusionAxion}
\end{figure}

\medskip

\noindent {\bf Search for a DM line below 2.3 keV.} 
To search for a DM line in this energy band, we add to the model described above (a power-law and instrumental lines) the absorbed X-ray emission of a thermal plasma and a power-law component, modeled through \textsc{XSPEC} as {\it TBABS*(APEC+Power-law)} . The plasma emission is modeled as {\it APEC}, while the {\it power-law} accounts for unresolved point sources or additional continuum emission. Both components are folded through the ARF and RMF response functions.
We perform the search fitting the 1–2.4 keV range to avoid edge effects and flux loss near the upper boundary. In this case, the full band is fitted simultaneously, rather than with sliding windows, to consistently model the astrophysical emission present in this interval, and because the size of the interval is rather small.

The free parameters of the astrophysical components described above are the plasma temperature and normalization for {\it APEC}~\footnote{The metal abundance is fixed to the solar value, using the abundances from~\cite{2000ApJ...542..914W}.}, the slope and normalization for the power-law, and the hydrogen column density for {\it TBABS}.
Only one instrumental line from the model of~\cite{EDRFWC} falls within this energy band, at approximately $1.49\text{--}1.50$ keV, with the exact centroid depending on the TM. Its energy is first determined through an initial fit and subsequently kept fixed for simplicity in the main analysis, following the same approach adopted for the instrumental line in the [2.3–9] keV band.

We remind that the absorption of the DM line is treated separately: it is modeled using {\it TBABS}, but the hydrogen column density is fixed to the value from observational data (HI4PI map). This choice ensures that signal normalization is not degenerate with background parameters and allows robust upper limits. The background model is intended only to provide a local description of the data rather than a precise characterization of the astrophysical emission.
By fixing DM absorption, we minimize the propagation of modeling uncertainties into the signal constraints without affecting the significance of a potential line detection.
As for the previous band, the background parameters are fitted independently for each TM. 
Since our goal is to search for a narrow DM line rather than to constrain the astrophysical background, this approach is conservative: it accounts for module-dependent calibration and instrumental differences, with the astrophysical parameters interpreted as an effective description rather than physical measurements.

\section{Results}
\label{Sec:Results}

The background model described in the previous section provides a satisfactory description of the data, as we show below. To quantify the goodness of fit, we combine the profile likelihoods in Eq.~\ref{eq:proflikelihood} of all seven TMs:
\begin{equation}
\tilde{C}(\phi_{\rm DM})
=
\min_{\{\vec{\Pi}_k\}}
\sum_k C^k(\phi_{\rm DM},\vec{\Pi}_k)
=
\sum_k \min_{\vec{\Pi}_k} C^k(\phi_{\rm DM},\vec{\Pi}_k),
\end{equation}
where the second equality holds because the nuisance parameters are independent across TMs.

We evaluate the background-only likelihood by setting $\phi_{\rm DM}=0$ and then derive the corresponding p-value to assess the goodness of fit of the background-only model. As mentioned above, we use the fact that the Eq.~\ref{eq:likelihood} asymptotically follows the $\chi^2$ distribution, to compute the associated p-value.
For the low energy band (1-2.4 keV) we find a p-value of 0.7. In the other energy band, the p-value depends on the specific energy window analyzed, and it is typically well above 0.05. 
The procedure described in Sec.~\ref{Sec:methodology} thus provides a satisfactory description of the background.

Then we scan for a possible DM line signal in energy steps smaller than the energy resolution of eROSITA and perform the line search described above.
For each energy, we quantify the significance of a possible DM line detection by computing $TS = \tilde{C}(\phi_{\rm DM=0})- \tilde{C}(\tilde{\phi}_{\rm DM})$, where $\tilde{\phi}_{\rm DM}$ is the DM line signal that minimizes $\tilde{C}$. 
Using the standard asymptotic one-sided profile-likelihood approximation, we interpret $\sqrt{\rm TS}$ as the local significance of a DM line signal.
We report this quantity in Fig.~\ref{fig:TS_scan}. 
Several local upward fluctuations are present, with the largest local significance exceeding $3 \, \sigma$, but none remains significant after accounting for the look-elsewhere effect. We therefore conclude that there is no statistically significant evidence for a DM line signal.
In general, the detection sensitivity is reduced in the presence of degeneracies between the line signal and the background model. This effect is expected to be most relevant at the energies of instrumental lines, whose locations are shown by the vertical dashed gray lines in Fig.~\ref{fig:TS_scan}. At these energies, the overlap with instrumental features reduces the local detection significance and results in weaker upper limits.

Finally, we derive 95\% confidence level (one-sided) upper limits on the DM line flux by requiring $
\Delta \tilde{C} = \tilde{C}(\phi_{\rm DM}) - \tilde{C}(\tilde{\phi}_{\rm DM}) = 2.71.$
The results are shown in the right panel of Fig.~\ref{fig:TS_scan}.
In Fig.~\ref{fig:DataWithModel} we show two examples of the DM line searches and the resulting upper limits in the two energy bands that we have considered. 
We translate the flux limits into constraints on the DM lifetime $\tau_{\rm DM}$ using
\begin{equation}
\label{eq:DMFlux}
  \phi_{\rm DM} = \frac{1}{4\pi\, m_{\rm DM}\, \tau_{\rm DM}} \, D \, N_\gamma ,
\qquad
D(\theta_0,\Delta\Omega) = 
\int_{\Delta\Omega} \mathrm{d}\Omega 
\int_{\rm l.o.s.} \mathrm{d}s \;
\rho\!\left(r[s,\theta,\theta_0]\right) .
\end{equation}

Here, $m_{\rm DM}$ denotes the mass of DM particles, which is equal to twice the line energy, for a two-body decay producing a monochromatic photon,  $E_\gamma = m_{\rm DM}/2$,
when the second final-state particle is massless or negligible in mass.
$N_\gamma$ denotes the number of photons produced per DM decay, and 
$D$ is the D-factor, defined as the integral of the DM density along the line of sight and over the solid angle $\Delta\Omega$. 
In our analysis, $\Delta\Omega$ corresponds to the $5^\circ$ circular region centered on the LMC, excluding the masked area. 
The quantity $\theta_0$ is the angle between the direction of observation and the halo center, $\theta$ is the angle between the line of sight and the pointing direction, $s$ denotes the distance along the line of sight, and $r$ is the distance to the halo center.

We model the DM density distribution in the LMC as a NFW profile~\cite{Navarro:1995iw}. We use the parameters derived in a recent analysis fitting Jeans dynamical model to Gaia DR3 stellar velocity measurements~\cite{2024A&A...692A..40K} (scale radius $r_s=19.9$ kpc and characteristic density $\rho_s=0.13$ GeV cm$^{-3}$, see Table I, second column).
The D-factor in the $5\degree$ radius circular region is $D=6.5\times10^{20}\,{\rm GeV cm}^{-2},$ while excluding the masked region it is $D=4.0\times10^{20}\,{\rm GeV cm}^{-2}.$

The lower limits on the DM lifetime are presented in Fig.~\ref{fig:ExclusionTau} for the case of one photon in the final state, i.e. $N_{\gamma}=1$. The gray area corresponds to the strongest upper limits from previous X-ray searches~\cite{Roach:2022lgo,Foster:2021ngm,Horiuchi:2013noa,Loewenstein:2008yi, Krivonos:2024yvm}.
The bounds derived in this work improve upon previous ones for $m_{\rm DM}\lesssim5$ keV, while they become progressively weaker for larger masses.

\begin{figure}[t]
   \centering
   \includegraphics[width=0.65\textwidth]{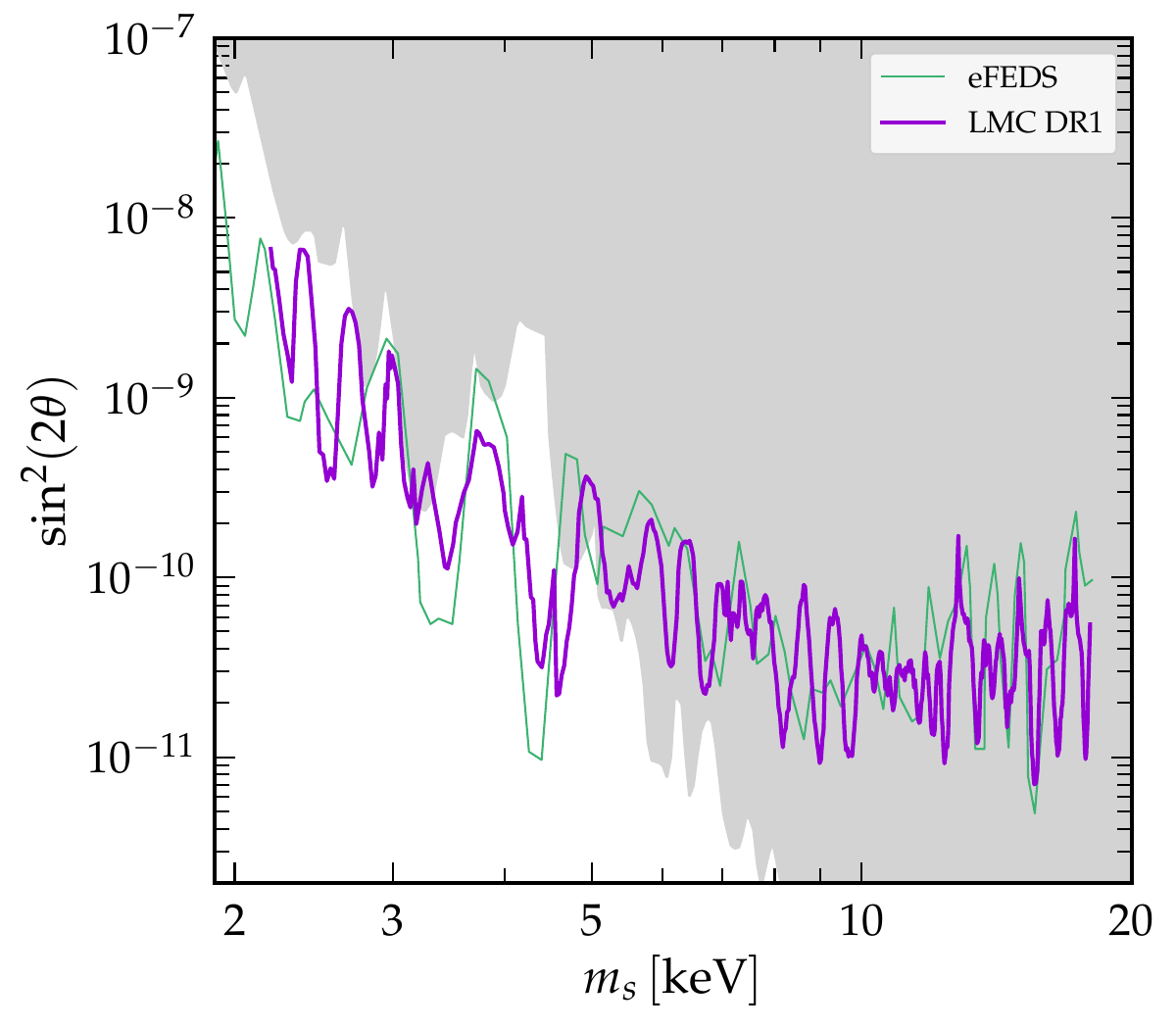}
   \centering
   \caption{Upper limits on the sterile neutrino mixing, $\sin^2\left(2\theta\right)$, as a function of the sterile neutrino DM mass, $m_s$. In purple we show the limits derived in this paper using LMC observations from eROSITA DR1. In green we show the limits coming from eROSITA eFEDS data~\cite{Fong:2024qeq}. In gray we show previously derived lower limits from other X-Ray observations including NuSTAR~\cite{Roach:2022lgo, Krivonos:2024yvm}, XMM-Newton~\cite{Foster:2021ngm}, Chandra~\cite{Horiuchi:2013noa} and Suzaku~\cite{Loewenstein:2008yi}.
   }
   \label{fig:ExclusionSterile}
\end{figure}

In Fig.~\ref{fig:ExclusionAxion} and Fig.~\ref{fig:ExclusionSterile} all the exclusion limits are recast in the parameter space of ALP and sterile neutrinos DM, using the decay rates in Eqs.~\ref{eq:rateALP} and~\ref{eq:rateSterile} to compute the lifetime.
In Fig.~\ref{fig:ExclusionSterile}, we also report the upper limits recently derived in~\cite{Fong:2024qeq} using the Final Equatorial Depth Survey (eFEDS) data from the eROSITA early data release. These limits are comparable to those derived in our analysis. 
Beyond these constraints, a lower bound on the mass of any fermionic DM candidate of $\mathcal{O}(1)$ keV is obtained from phase-space considerations based on observations of galaxies~\cite{Boyarsky:2008ju,Gorbunov:2008ka,Alvey:2020xsk,Bezrukov:2025ttd}. For the case of sterile neutrino DM, stronger limits, dependent on the specific production mechanism, can also be obtained from galaxy counts, Lyman-$\alpha$ observations, and Big Bang nucleosynthesis (BBN), see~\cite{Cherry:2017dwu,Wang:2017hof,DES:2020fxi,Dekker:2021scf,Lovell:2015psz,Zelko:2022tgf,Newton:2024jsy,Shi:1993hm,Bodeker:2020hbo,Akita:2025txo, Gorbunov:2025nqs} and references therein.

Finally, in the right panel of Fig.~\ref{fig:ExclusionAxion}, we interpret our results in the context of the irreducible axion background studied in~\cite{Langhoff:2022bij}. In this scenario, ALPs do not constitute the entirety of the DM density. Instead, a subdominant population is unavoidably produced via freeze-in through particle scatterings in the primordial plasma of the early Universe. Ref.~\cite{Langhoff:2022bij} assumes a minimal ALP abundance corresponding to a reheating temperature of $T_{\rm RH}=5$ MeV, which is the lower limit allowed by BBN. Under this assumption, we derive the limits shown in Fig.~\ref{fig:ExclusionAxion}. Note that our constraints vanish at large couplings, since in this regime fast ALP decays exponentially suppress the ALP density.

\section{Conclusions}
\label{Sec:Conclusion}

In this work, we have exploited the first eROSITA-DE data release to search for X-ray line emission from decaying DM. We have analyzed observations of the Large Magellanic Cloud, which, due to its large mass and its relative closeness, constitutes an excellent target for DM searches.

We have not found any statistically significant evidence for a DM X-ray line signal and we have derived upper limits on the DM lifetime, interpreting them within the parameter spaces of sterile neutrino and ALP DM models. Our constraints are especially stringent for DM masses below $\sim 5$ keV, exceeding those obtained with other X-ray telescopes.

Exploiting the large-sky survey strategy of eROSITA, further improvements in sensitivity can be expected from analyses of larger fractions of the observed sky~\cite{Dekker:2021bos,Fong:2024qeq}, and/or by combining the spectral search with a morphological analysis that exploits the expected spatial distribution of DM in the target region. In particular, all-sky searches for DM line emission and cross-correlation studies combining eROSITA X-ray data with galaxy surveys~\cite{Caputo:2019djj} offer promising avenues to enhance the discovery potential. We plan to explore these directions in future works. Finally, a second eROSITA data release (DR2) is expected in the near future~\footnote{\href{https://erosita.mpe.mpg.de/erass/}{https://erosita.mpe.mpg.de/erass/}}.

\section*{Acknowledgements}
We thank Miltiadis Michailidis for helpful discussions on the eROSITA data and software. JTC thanks Enzo A. Saavedra and Federico García for useful discussions on XSPEC usage.

JTC, MR and  MT acknowledge support from the  Research grant TAsP (Theoretical Astroparticle Physics) funded by \textsc{infn}. The work of JTC and MT is supported by the Italian Ministry of University and Research (MIUR) via the PRIN 2022 Project No. 2022F2843 “Addressing systematic uncertainties in searches for dark matter” funded by MIUR.
JTC and AC acknowledge the support by the European Research Area (ERA) via the UNDARK project (project number 101159929). AC is supported by an ERC STG grant (``AstroDarkLS'', grant No. 101117510). AC acknowledges the Weizmann Institute of Science for hospitality at different  stages of this project and the support from the Benoziyo Endowment Fund for the Advancement of Science. The work of MR is supported by the European Union – Next Generation EU and by the Italian Ministry of University and Research (MUR) via the PRIN 2022 project n. 20228WHTYC - CUP D53C24003550006

This work is based on data from eROSITA, the soft X-ray instrument aboard SRG, a joint Russian-German science mission supported by the Russian Space Agency (Roskosmos), in the interests of the Russian Academy of Sciences represented by its Space Research Institute (IKI), and the Deutsches Zentrum für Luft- und Raumfahrt (DLR). The SRG spacecraft was built by Lavochkin Association (NPOL) and its subcontractors, and is operated by NPOL with support from the Max Planck Institute for Extraterrestrial Physics (MPE). The development and construction of the eROSITA X-ray instrument was led by MPE, with contributions from the Dr. Karl Remeis Observatory Bamberg \& ECAP (FAU Erlangen-Nuernberg), the University of Hamburg Observatory, the Leibniz Institute for Astrophysics Potsdam (AIP), and the Institute for Astronomy and Astrophysics of the University of Tübingen, with the support of DLR and the Max Planck Society. The Argelander Institute for Astronomy of the University of Bonn and the Ludwig Maximilians Universität Munich also participated in the science preparation for eROSITA.  The eROSITA data shown here were processed using the eSASS software system developed by the German eROSITA consortium.

\bibliographystyle{bibi}
\bibliography{biblio.bib}

\end{document}